\documentclass[twocolumn,prd,amssymb,showpacs,10pt]{revtex4}
\usepackage{graphicx}% Include figure files

\usepackage{bm}% bold math

\def\l4{l_{\rm 4}}
\def\planck5{l_{\rm 5}}

\def\AdS5{\rm AdS_5}

\begin{document}

\title{Cosmology and the Korteweg-de Vries Equation}
\author{James E. Lidsey}
\affiliation{Astronomy Unit, School of Physics and Astronomy,   
Queen Mary University of London, Mile End Road, LONDON, E1 4NS, U.K. \\  
J.E.Lidsey@qmul.ac.uk}

%\date{\today}

\begin{abstract}
The Korteweg-de Vries (KdV) equation is a non-linear 
wave equation that has played a fundamental role
in diverse branches of mathematical and theoretical  
physics. In the present paper, we consider its significance 
to cosmology. It is found that 
the KdV equation arises in a number of important scenarios, including 
inflationary cosmology, the cyclic universe, loop quantum cosmology 
and braneworld models.
Analogies can be drawn between cosmic dynamics and the propagation 
of the solitonic wave solution to the equation, 
whereby quantities such as the 
speed and amplitude profile of the wave can be identified with cosmological 
parameters such as the spectral index of the density perturbation spectrum and
the energy density of the universe. The unique mathematical
properties of the Schwarzian derivative operator are important to the 
analysis. A connection with dark solitons in Bose-Einstein condensates 
is briefly discussed.  
\end{abstract}

\vskip 1pc \pacs{98.80.Cq}
\maketitle

\section{Introduction}

\def\theequation{\arabic{equation}}

The Korteweg-de Vries (KdV) equation \cite{KDV} is the completely integrable, 
third-order, non-linear partial differential equation (PDE):  
\begin{equation}
\label{kdv}
\partial_t u + \partial^3_x u + \frac{3}{u_0} u \partial_x u =0 ,
\end{equation}
where $u= u (x,t)$, $\partial_t= \partial /\partial t$, 
$\partial^3_x= \partial^3/\partial x^3$, etc., $u_0$ is a constant
and $(x,t)$ represent space and time coordinates, respectively. 
This equation was originally derived within the context of 
small-amplitude, non-linear water wave theory and 
it is well known that it admits a solitonic wave solution 
of the form 
\begin{equation}
\label{kdvsoliton}
u=u_0\lambda^2 {\rm sech}^2 [\lambda (x- \lambda^2 t)/2] ,
\end{equation}
where the constant $\lambda/2$ represents the wavenumber of the soliton 
\cite{KDV}. 
The KdV soliton is characterized by the property that its speed and 
amplitude are proportional to the square of the wavenumber. 

The KdV equation has played a central role in diverse 
branches of physics, including nonlinear optics, atomic and nuclear physics, 
Bose-Einstein condensates and astrophysical plasmas (see Ref. \cite{scott}
for a review.) As far as we are aware, however,  
it has not been discussed previously within  
a cosmological context (although see Ref. \cite{rosu}). 
This is perhaps not surprising, given that 
the KdV equation is a third-order 
PDE in two independent variables, whereas the 
field equations for spatially isotropic universes are second-order, 
ordinary differential equations (ODEs). 

One the other hand, Eq. (\ref{kdv}) can be reduced to the non-linear ODE:  
\begin{equation}
\label{kdvphi}
- \lambda^2 u' + u''' + \frac{3}{u_0} u u' =0 ,
\end{equation}
where a prime denotes $d/d\phi$ and $\phi \equiv x - \lambda^2 t$
represents a `wave-like' independent variable. 
The purpose of the present paper is to show that 
wave solutions to Eq. (\ref{kdvphi}), and  
in particular the soliton solution (\ref{kdvsoliton}), arise in a number 
of cosmological settings, including the inflationary paradigm, the 
cyclic universe scenario, loop quantum cosmology and braneworld models. 
We consider scenarios where the universe is dominated by a single,
self-interacting scalar field, $\phi$. Interpreting the value of the 
scalar field in terms of a wave-like coordinate of the KdV equation 
then allows for a direct analogy 
to be drawn between cosmological dynamics 
on the one hand and solitonic behaviour on the other. 
Cosmological parameters can then be identified with 
quantities such as the speed and amplitude of the 
corresponding wave. 

The paper is organised as follows. In Section 2, we discuss a 
connection between the KdV equation and the Schwarzian derivative 
operator that we employ in later Sections. 
We proceed in Section 3 to consider the classes of 
inflationary and cyclic universes
that generate density perturbation spectra with a  
constant spectral index. In Section 4, we find that the scaling solutions 
of various braneworld and loop quantum cosmological scenarios are 
analogous to the 
KdV soliton. We conclude with a discussion in 
Section 5 on a connection with dark solitons in Bose-Einstein 
condensates. Unless otherwise stated, units 
are chosen such that $\hbar = c =1$ and the Planck mass is normalised to
$m_{\rm P} =\sqrt{8\pi}$. 

\section{The KdV Equation and the Schwarzian Derivative Operator}

\def\theequation{\arabic{equation}}

The KdV equation (\ref{kdvphi}) admits an 
auto-B\"acklund transformation, whereby
a solution $u=u(\phi)$ can be derived
from a given seed solution $\bar{u} = \bar{u}(\phi)$ \cite{lax}.
For the special case where the seed is the trivial solution 
$\bar{u}=0$, such a transformation 
reduces to the condition that a solution to Eq. (\ref{kdvphi}) is given by  
\begin{equation}
\label{kdvsol}
u = u_0 (\lambda^2 -4y^2) ,
\end{equation}
where $\lambda^2$ is a constant and the function $y=y(\phi )$ is a 
particular solution to
the first-order Riccati equation: 
\begin{equation}
\label{yriccati} 
y' = \frac{u}{4u_0} = \frac{\lambda^2}{4} - y^2 .
\end{equation}
For example, the solution 
\begin{equation}
\label{ytanh}
y=\frac{\lambda}{2} \, {\rm tanh} (\lambda \phi /2)
\end{equation}
generates the KdV soliton (\ref{kdvsoliton}). 

Solutions to Eq. (\ref{yriccati}) satisfy a third-order 
ODE given by
\begin{equation}
\label{schy}
S[y(\phi)] \equiv \frac{y'''}{y'} - \frac{3}{2} \left( 
\frac{y''}{y'} \right)^2 = - \frac{\lambda^2}{2} ,
\end{equation}
as may be verified by direct differentiation of Eq. (\ref{yriccati}). 
The left-hand side of Eq. (\ref{schy}) is the Schwarzian derivative 
operator (often referred to simply as the {\em Schwarzian}) \cite{hille}. 
This operator exhibits a number of remarkable properties, 
one of which we exploit in the present
work: it is the unique combination of derivatives that is invariant 
under a homographic transformation corresponding to the group of 
fractional linear transformations. This follows since, for a function
$y=y(\phi)$, the composition 
$N(y) \equiv [\ln (y')]'=y''/y'$ transforms under the inversion $y \rightarrow
1/y$ such that $N(1/y)= N(y) -2 y'/y$. Some straightforward 
algebra then implies that the operator $S(y) \equiv 
N' -N^2/2$ is invariant under
inversion. Moreover, since $S(y)=S(my+n)$ for any $m,n\in \Re$, 
$S(y)$ is invariant under the full group of fractional linear transformations.

This implies that if $\overline{y}(\phi )$ is a particular `seed' 
solution to the differential equation  
$S[y] = f(\phi)$ for some function $f(\phi)$, the {\em general solution}
to such an equation is given by 
\begin{equation}
\label{gensolW}
y(\phi) = \frac{a \overline{y} +b}{c\overline{y}+d}, \qquad 
\left( 
\begin{array}{cc}
a &  b \\
c & d
\end{array}
\right) \in {\rm SL(2,R)} \, ,
\end{equation}
where $a,b,c,d$ are constants such that $ad-bc =1$. 

To summarize, therefore, if a particular solution to 
the Schwarzian equation (\ref{schy}) for constant $\lambda$ 
can be found, the general solution (\ref{gensolW}) 
can be written down immediately. 
Restricting the general solution to satisfy the first-order constraint
(\ref{yriccati}) then generates a solution (\ref{kdvsol}) to the KdV equation
(\ref{kdvphi}). We employ this result in the following Sections 
for a number of cosmological models. 

\section{Cosmological Density Perturbations and the KdV Soliton}

\def\theequation{\arabic{equation}}

\subsection{The Inflationary Universe}

The inflationary scenario remains the cornerstone 
of modern, early universe cosmology. 
Whilst there is currently considerable interest in 
multiple-field versions of the paradigm, our aim in this Section is to revisit 
the simplest version of the scenario, namely inflation 
driven by a single, minimally coupled, 
slowly-rolling scalar inflaton field, $\phi$. (For reviews, see, e.g., 
Ref. \cite{lidseyreview,lyth}.) We consider the 
general class of models that generate a {\em scale-invariant} 
density perturbation spectral index (hereafter referred to as the spectral
index) to lowest-order in the slow-roll approximation. 
Our main aim is to highlight some interesting mathematical features of 
the underlying differential equations that have not been 
previously discussed. 

The cosmological Friedmann equations in Hamilton-Jacobi form are given by
\begin{equation}
\label{friedmann}
3H^2  -2H'^2= V (\phi) , \qquad 
\dot{\phi} =-2H' ,
\end{equation}
where the Hubble parameter $H=H(\phi)$ is viewed as a function of $\phi$,
$V(\phi)$ denotes the inflaton potential and 
a prime and dot denote differentiation with respect to the inflaton 
field and cosmic time, respectively. 
The energy density of the universe is 
$\rho (\phi )= 3 H^2(\phi)$. It is assumed implicitly 
and without loss of generality that the inflaton
varies monotonically with cosmic time such that $\dot{\phi} >0$ 
($H' <0$). 

The Hubble slow-roll parameters are defined by
\begin{equation}
\label{defepsilon}
\epsilon \equiv 2 \frac{H'^2}{H^2} , \qquad 
\eta \equiv  2 \frac{H''}{H} ,
\end{equation}
and the spectral index is given by 
\begin{equation}
\label{spectralindex}
1-n_s = \frac{2}{(1-\epsilon)^2} \left[ \epsilon - 
\frac{(1-\epsilon^2)}{2} \left( \frac{d \ln \epsilon}{d {\cal{N}}} \right) 
\right] \, ,
\end{equation}
where higher-order derivative terms in 
${\cal{N}} \equiv - \ln (aH)$ have been neglected.  
To lowest-order in the slow-roll approximation, 
$1-n_s= 4\epsilon -2 \eta$. This condition 
is a second-order, non-linear ODE for the 
dependent variable $H=H(\phi)$: 
\begin{equation}
\label{ns}
4 \frac{H''}{H} -8 \frac{H'^2}{H^2} = -( 1-n_s) .
\end{equation}

Considerable insight 
into the nature of the general solution to Eq. (\ref{ns}) may be gained by 
expressing the energy density of the universe in terms of the gradient 
of a `potential' function, $W=W(\phi)$, such that 
\begin{equation}
\label{defW}
H^2 (\phi) \equiv  4 H_0^2W'(\phi) ,
\end{equation}
where $H_0$ is an arbitrary constant. 
Substituting this definition into Eq. (\ref{ns}) yields
\begin{equation}
\label{schW}
S[W(\phi )] \equiv 
\frac{W'''}{W'} - \frac{3}{2} \left( \frac{W''}{W'} \right)^2 
= - \frac{\lambda^2}{2} ,
\end{equation}
where $\lambda^2 \equiv 1- n_s$. 

The left-hand side of Eq. (\ref{schW}) is the Schwarzian derivative 
of the function $W(\phi)$ and it is interesting that it arises 
in this cosmological context. Consequently, 
we may now determine from Eqs. (\ref{defW}) and (\ref{schW}) 
the general forms of the Hubble
parameters for the full family of inflationary 
cosmologies that generate a constant spectral index.  

Since current observational bounds on the spectral index 
inferred from the WMAP7$+$H0 data set are $0.939<n_s<0.987$
at the $2\sigma$ confidence limit \cite{komatsu}, we focus on the 
red perturbation spectrum, $\lambda^2 >0$.
A particular solution to Eq. (\ref{schW})
is $\overline{W}(\phi) = \exp(\lambda \phi)$ and 
the general solution is therefore given directly by 
\begin{eqnarray}
\nonumber
W(\phi ) & = & \frac{a e^{\lambda \phi} +b}{c e^{\lambda \phi} +d}
\\ 
\label{genred}
H^2(\phi) & = & 4 \lambda H_0^2 \frac{e^{\lambda \phi}}{\left( 
c e^{\lambda \phi} +d \right)^2} .
\end{eqnarray}

Moreover, comparing Eq. (\ref{schW}) 
with Eq. (\ref{schy}) and Eq. (\ref{defW}) with Eq. (\ref{yriccati})
immediately implies that $H^2 (\phi)$ satisfies the KdV equation 
\begin{equation}
\label{kdvcosmic}
-(1-n_s) {H^2}^{\prime} +{H^2}^{\prime \prime \prime} 
+\frac{3}{H_0^2} H^2 {H^2}^{\prime} =0
\end{equation}
{\em if} the general solution to Eq. (\ref{schW}) is restricted to satisfy 
the Riccati equation 
\begin{equation}
\label{ifW}
W' = \frac{\lambda^2}{4} - W^2 .
\end{equation}

It may be verified 
that condition (\ref{ifW}) is satisfied if  
\begin{equation}
\label{iff}
ad=-bc =1/2, \qquad \lambda = \frac{1}{cd} = \frac{2a}{c} .
\end{equation}
As a result, the solution (\ref{genred}) can be expressed as  
\begin{equation}
\label{Hsoliton}
H^2(\phi)  =  H^2_0 \, \lambda^2 \, 
{\rm sech}^2 \left( \frac{\lambda}{2} \frac{\sqrt{8\pi}}{m_{\rm P}}
\phi \right)
\end{equation}
for $cd  >0$ and 
\begin{equation}
\label{cosechsol}
H^2(\phi)  =  - H^2_0 \, \lambda^2 \, 
{\rm cosech}^2 \left( \frac{\lambda}{2} \frac{\sqrt{8\pi}}{m_{\rm P}}
\phi \right) 
\end{equation}
if $cd   <0$, 
where we have specified $c=|d|$ without loss of generality 
and have restored the dependence on the Planck mass for future reference.
(If $c \ne |d|$, Eqs. (\ref{Hsoliton})-(\ref{cosechsol}) can be recovered 
by performing a linear translation $\lambda^{-1} \ln | d/c|$ on the 
value of the inflaton field.)  

Given the general form of the Hubble parameter, the 
inflationary potentials can be deduced directly 
from the Hamilton-Jacobi  
equation (\ref{friedmann}). It is straightforward to verify 
that these potentials correspond precisely to those derived in \cite{vall}
to lowest-order in 
slow-roll. From the cosmic dynamical systems point of view, 
the late-time attractor of these models is the power-law solution 
$\epsilon = \lambda^2$. There are two different 
potentials depending on whether the initial value of $\epsilon$ is 
greater or less than $\lambda^2$ \cite{vall}. Within the context of the 
present discussion, the power-law model is the seed solution 
$\overline{W}(\phi) = \exp(\lambda \phi)$ and the two models are 
characterized by ${\rm sgn} (cd)$, i.e., by ${\rm sgn} (W')$.  

Eqs. (\ref{Hsoliton}) and (\ref{cosechsol}) are both wave solutions 
to the KdV equation (\ref{kdvcosmic}) and the former 
has precisely the form of the non-singular
KdV soliton (\ref{kdvsoliton}). This suggests a direct analogy 
can be drawn between such a wave and inflationary cosmology. 
In such an analogy,  
the  inflaton field plays the role of a characteristic, wave-like coordinate
on a two-dimensional spacetime $\{ x ,t\}$, 
the speed of the soliton is determined by the deviation of the spectral 
index away from the scale-invariant, Harrison-Zel'dovich
spectrum $n_s =1$, and the amplitude profile of the soliton 
is parametrized by the energy density of the universe (in appropriate 
units).  

The first slow-roll parameter for the `soliton' solution (\ref{Hsoliton}), 
is given by   
\begin{equation}
\label{epsilonsol}
\epsilon (\phi) = \frac{\lambda^2}{2} {\rm tanh}^2 \left( 
\lambda {\phi} /2 \right)
\end{equation}
and determines the tensor-scalar 
ratio, $r \equiv {\cal{P}}_T^2 / {\cal{P}}_S^2 = 16 \epsilon$, 
where ${\cal{P}}_S^2 =H^2/(64 \pi^4 \epsilon )$ and 
${\cal{P}}_T^2 =H^2/(4\pi^4 m_{\rm P}^2)$ are the amplitudes of the 
density and gravitational wave perturbations \cite{lidseyreview,lyth}. 
This parameter is bounded from above
such that $r < 8(1-n_s)$, which is 
consistent with the current observational 2$\sigma$ bound $r<0.24$
\cite{komatsu}. 

\subsection{The Cyclic Universe}

During inflation, quantum fluctuations in the inflaton field 
become frozen on super-Hubble radius scales because 
the comoving Hubble scale decreases with time 
due to the rapid, accelerated expansion of the universe. 
However, the comoving Hubble radius 
can also decrease if the universe undergoes a phase of slow, decelerated 
{\em contraction} driven by a negative scalar field 
potential. This is the basis of the 
cyclic universe scenario. As shown in \cite{cyclic}, the expression 
(\ref{spectralindex}) for the spectral index 
is invariant under the duality $\epsilon 
\rightarrow 1/\epsilon$ and this implies there exists a one-to-one 
correspondence between inflationary and cyclic models that generate identical 
spectral indices. 

This suggests that a similar gravity/soliton analogy
may be established for the cyclic universe scenario. 
Indeed, in the Hamilton-Jacobi formalism of the cosmological Friedmann 
equations (\ref{friedmann}), 
the definition of the Hubble parameter $H = \dot{a}/a$ in terms of 
the scale factor $a$ implies that 
\begin{equation}
\label{defH}
a'H' =-\frac{1}{2} a H .
\end{equation}
Integrating Eq. (\ref{defH}) then yields the 
dependence of the scale factor on the scalar field 
in terms of the quadrature
\begin{equation}
\label{quadscale}
a (\phi) = a_0 \exp \left[ -\frac{1}{2} \int^{\phi} d \phi 
\frac{H}{H'} \right] ,
\end{equation}
where $a_0$ is an arbitrary constant.

However, Eq. (\ref{defH}) is invariant under the
simultaneous interchange $H(\phi) \leftrightarrow a(\phi)$. If we 
therefore consider 
a `dual' cosmology where the Hubble parameter is given by 
$\tilde{H}(\phi) = a(\phi)$, Eq. (\ref{quadscale}) implies that the
dual scale factor is given by the quadrature
\begin{equation}
\label{newscale}
\tilde{a}(\phi) = \tilde{a}_0 \exp \left[ -\frac{1}{2} \int^{\phi} 
d\phi \frac{a}{a'} \right] .
\end{equation}
Since the seed cosmology $\{ H(\phi), a(\phi) \}$ itself
satisfies Eq. (\ref{defH}), the duality between the two scenarios 
is given by the simultaneous interchange of the Hubble parameters and 
scale factors of the two scenarios when all parameters are expressed as 
functions of the scalar field \cite{triality}: 
\begin{equation}
\label{duality}
\tilde{a}(\phi ) = H(\phi) , \qquad \tilde{H}(\phi) = a (\phi) .
\end{equation}
It is straightforward to verify that
under this duality, the Hubble slow-roll parameter 
indeed transforms as $\tilde{\epsilon} = 1/ \epsilon$.

As a result of this duality, the analysis of Section 3.1 applies directly to
the cyclic universe. We conclude, therefore, that 
there exists a one-to-one correspondence 
between solutions to the KdV equation and the respective 
cyclic cosmological models when the spectral index is constant. 
For the cyclic model that is dual to Eq. (\ref{Hsoliton}), the speed of the
soliton is once more determined by the spectral index, 
whereas the amplitude of the soliton is now proportional 
to the square of the cosmic scale factor.   

\section{Modified Cosmology}

In recent years, considerable interest has focused on cosmological
dynamics arising from modified gravity theories. This is motivated in part
by open questions in early universe cosmology, such as the singularity
problem, as well as providing alternative scenarios to dark energy models. 
At a phenomenological level, such modifications can be quantified 
by altering the standard form of the Friedmann equation such that 
\begin{equation}
\label{modfried}
3 H^2 = \rho L^2 (\rho ) ,
\end{equation}
where $L=L(\rho )$ is a given function of the energy density 
and is determined by the specific model in question. 

Of particular interest in such scenarios are scaling (attractor) solutions, 
since these enable the generic asymptotic behaviour of a 
cosmological model to be better understood. Scaling solutions 
are characterized by the property that the energy densities of the component 
matter fields scale at the  same rate as the universe expands (contracts). 
Such solutions were classified in Ref. \cite{lidseyscale} under the assumption 
that the matter content of the universe is comprised of 
a self-interacting scalar field with potential $V(\phi)$ 
and a barotropic fluid with an adiabatic index $\gamma$. 
By introducing a parameter
\begin{equation}
\label{deflambda}
\lambda \equiv -\frac{1}{L} \frac{V'}{V}
\end{equation}
it was found that $\lambda = {\rm constant}$ 
is a necessary condition for a scaling 
solution to exist. In that case, there exists an attractor solution 
for $\lambda^2 > 3 \gamma$, where the  relative contribution of the 
scalar field energy density to the total density of the universe is 
$\Omega_{\phi} = 3\gamma/\lambda^2$. There exists a second stable solution 
if $\lambda^2 <6$ where the scalar field dominates the fluid. 
It was further shown that these solutions exist if the condition 
\begin{equation}
\label{scalingcon}
\rho \frac{\rho^{\prime \prime}}{\rho^{\prime 2}} -1 
= \rho \frac{d \ln [L(\rho)]}{d \rho}
\end{equation}
is satisfied \cite{lidseyscale}. 

In this Section, we investigate the conditions that allow 
for scaling solutions in modified gravity to be interpreted as 
KdV-type solitons. We proceed by rewriting Eq. (\ref{scalingcon}) in the form
\begin{equation}
\label{scalecon1}
\frac{\rho''}{\rho} - \frac{3}{2}  \frac{\rho'^2}{\rho^2} = 
\frac{\rho'^2}{\rho^2} \left[ \rho \frac{d \ln L}{d \rho} - 
\frac{1}{2} \right]  .
\end{equation}
Defining a new 
variable $\sigma = \sigma (\phi)$: 
\begin{equation}
\label{defsigma}
\rho = 4 \rho_0 \sigma'  ,
\end{equation}
where $\rho_0$ is an arbitrary constant, then transforms 
Eq. (\ref{scalecon1}) into a Schwarzian differential equation: 
\begin{equation}
\label{schscale}
S[\sigma (\phi)] = \left( \frac{\sigma''}{\sigma'} \right)^2 
\left[ \rho \frac{d \ln L}{d \rho} - 
\frac{1}{2} \right]  ,
\end{equation}
where the dependence of the square bracket on $\sigma$ is implicit. 

We now look for particular solutions to Eq. (\ref{schscale}) 
that satisfy the condition that the Schwarzian of $\sigma$ is constant: 
\begin{equation}
\label{constantsch}
S[\sigma (\phi) ] = - \frac{\lambda^2}{2}  .
\end{equation}
The discussion of Section 
3 implies that the energy density of the universe 
will satisfy the KdV equation if $\sigma' = (\lambda^2/4) -\sigma^2$. 
As we saw above, the are two possible solutions when $\lambda^2 >0$, depending
on whether $\sigma' >0$ or $\sigma'<0$. (The product $\rho_0 \sigma'$ is 
assumed implicitly to be positive-definite.)
 
Eqs. (\ref{schscale}) and (\ref{constantsch}) imply that   
\begin{equation}
\label{existsol}
\left( \frac{\sigma''}{\sigma'} \right)^2 
\left[ \rho \frac{d \ln L}{d \rho} - 
\frac{1}{2} \right] = -\frac{\lambda^2}{2}  .
\end{equation}
Since $\sigma''= -2\sigma \sigma'$, it follows that 
\begin{equation}
\left( \frac{\sigma''}{\sigma'} \right)^2 = \lambda^2 -
\frac{\rho}{\rho_0}  .
\end{equation}
Consequently, Eq. (\ref{existsol}) simplifies to the first-order ODE 
\begin{equation}
\label{Lequation}
\frac{d \ln L}{d \rho} = -\frac{1}{2} \frac{1}{\rho_0 \lambda^2 - \rho}
\end{equation}
and integrating yields the solution 
\begin{equation}
\label{Lsolution}
L =\left( 1- \frac{\rho}{\rho_0 \lambda^2} \right)^{1/2}  ,
\end{equation}
where we have chosen the appropriate branch of the general solution and the 
constant of integration to ensure that the standard relativistic cosmology 
is recovered in the low-energy limit $\rho \ll \rho_0 \lambda^2$. 
The modified Friedmann equation (\ref{modfried}) is 
therefore given by 
\begin{equation}
\label{kdvfriedmann}
3 H^2 =  \rho \left( 1- \frac{\rho}{\rho_0 \lambda^2} \right)  .
\end{equation}
Finally, the results of Section 3 can be carried over to deduce that 
the corresponding scaling solutions, 
when expressed in terms of the energy density, are given by the 
KdV wave-solutions 
\begin{equation}
\label{lqcscale}
\rho = \rho_0 \lambda^2 {\rm sech}^2 (\lambda \phi /2) 
\end{equation}
when $\sigma' >0$ ($\rho_0 >0$) and 
\begin{equation}
\label{branescale}
\rho = -\rho_0 \lambda^2 {\rm cosech}^2 (\lambda \phi /2) 
\end{equation}
when $\sigma' <0$ ($\rho_0 <0$). 

The Friedmann equation (\ref{kdvfriedmann}) arises in a number of 
cosmological models that are directly
motivated by quantum gravity considerations. 
When $\rho_0 < 0$, the model corresponds to the Randall-Sundrum braneworld
scenario \cite{rs}, where our observable universe 
is interpreted as a co-dimension one-brane 
embedded in five-dimensional, ${\rm Z}_2$ symmetric 
anti-de Sitter space. The coefficient $\rho_0 \lambda^2$ is determined 
by the tension of the brane \cite{rs}. On the other hand, the case where 
$\rho_0 >0$  results in the Friedmann equation for the Shtanov-Sahni (S-S)
bouncing braneworld \cite{sahni}. 
In this model, the universe is again interpreted as 
a one-brane embedded in a five-dimensional spacetime sourced by a
bulk cosmological constant, but the extra fifth dimension is now
assumed to be timelike. 

Furthermore, the Friedmann equation (\ref{kdvfriedmann})  
arises generically in loop quantum cosmological (LQC) scenarios when 
$\rho_0 >0$ \cite{param}. 
In LQC models, the quadratic corrections 
in the energy density originate from 
non-perturbative quantum geometric corrections and become important at high
energy scales. Indeed, in such a framework, $\rho_0 \lambda^2$ 
determines a critical density
\begin{equation}
\label{rhocrit}
\rho_0 \lambda^2 = \rho_{\rm crit} = \frac{\sqrt{3}}{16\pi^2 \gamma^3}
\rho_{\rm Pl} , 
\end{equation}
where $\rho_{\rm Pl}$ is the Planck density and $\gamma \approx 0.2375$
is the Immirzi parameter \cite{param}.

In effect, therefore, by focusing on the KdV equation we have arrived at 
three different cosmological models 
that are all inspired by quantum gravity effects.
The corresponding scaling 
solutions (\ref{lqcscale}) and (\ref{branescale}) are those found 
previously in \cite{lidseyscale,lidseyhawk}. We may now interpret these
solutions as wave solutions of the KdV equation. The non-singular soliton
solution reflects the non-singular nature of the S-S and LQC  
scaling solutions. 
Due to the nature of the quantum corrections,
the universe collapses from infinity $(\phi 
\rightarrow -\infty$ ), undergoes a non-singular bounce at $\rho = 
\rho_{\rm crit}$ $(\phi =0) $, and then expands to infinity $(\phi \rightarrow 
+\infty )$. From the point of view of a stationary laboratory `observer', 
such dynamics would be equivalent to the 
time-dependence of the soliton amplitude as the wave propagates  
(modulo the appropriate 
relationship between cosmic and laboratory times). The cosmic bounce
corresponds to the passing of the peak of the wave. 
In the LQC scenario, the `speed' of the wave 
is parametrised by the fractional energy density of 
the scalar field and the barotropic index of the fluid, 
$\lambda^2 = 3 \gamma / \Omega_{\phi}$. 

\section{Discussion}

\def\theequation{\arabic{equation}}

In the present work, it has been shown for the first time 
that the KdV equation 
arises in a number of important cosmological scenarios, including 
the inflationary universe, the cyclic universe, loop quantum cosmology
and braneworld cosmology. In each model, cosmological solutions can be 
reinterpreted as wave-like 
solutions to the KdV equation, and this allows for analogies to
be drawn between cosmic dynamics and wave propagation.  

For example, in the inflationary scenario, we have found that the ODE 
determining the spectral index of the density perturbation spectrum 
generated during single-field, 
slow-roll inflation is closely connected to both the
Schwarzian derivative operator and the KdV wave equation. In principle,  
this allows for the full family of inflationary models that generate a 
constant spectral index to be classified in a very straightforward 
manner in terms of solutions to the KdV equation. 

In the region of observational 
parameter space $r < 8(1-n_s)$, a formal analogy 
was established between the non-singular KdV soliton and the 
inflating universe. 
In such a correspondence, the scalar field plays the role of a 
wave-like coordinate, the speed of the 
soliton is determined by the value of the spectral index
and the amplitude of
the soliton is parametrized by the energy density of the universe. 
Due to the inflationary/cyclic 
universe duality, similar conclusions 
hold for the simplest version of the cyclic universe scenario, although in this 
case the amplitude of the soliton is related to the scale factor of the 
universe.

The general conditions for scaling solutions in a 
class of modified cosmological models sourced by a scalar field 
and a perfect fluid were considered. Requiring that the integral of the
cosmic energy density (with respect to the scalar field) has a
constant Schwarzian derivative led naturally to a modified Friedmann equation
that arises generically in the Randall-Sundrum and 
Shtanov-Sahni 
braneworld models \cite{rs,sahni} and loop quantum cosmology scenarios
\cite{param}. In all cases, the corrections 
to the Friedmann equation are quadratic in the energy density. For such models, 
the scaling solutions may be interpreted as wave solutions to the 
KdV equation, where the 
cosmic energy density is again analogous to the soliton wave amplitude. 

Finally, it is worth remarking 
that the KdV equation is closely related to the 
non-linear Schr\"odinger (NLS) equation. This equation 
admits solutions that determine the propagation of solitons in Bose-Einstein
condensates \cite{tsu,zak}.  
A Bose-Einstein condensate is the ground state of a gas of $N$ interacting 
bosons trapped by an external potential. In the limit where the interaction 
between the atoms is sufficiently weak, the mean-field approximation may 
be employed. In this case, the macroscopic wavefunction for the condensate, 
$\psi$, satisfies the Gross-Pitaevskii (GP) equation \cite{gp}:
\begin{equation}
\label{GP}
i \hbar \partial_t \psi = - \frac{\hbar^2}{2m} \nabla^2 \psi + 
V({\mathbf{x}}, t) \psi + g | \psi |^2 \psi ,
\end{equation}
where $V({\mathbf{x}},t)$ is the trapping potential 
and $m$ is the mass of the atoms forming the condensate. The scattering 
coefficient is given by $g=4 \pi \hbar^2 N a/m$, where $a$ is the 
(s-wave) scattering length. 

By employing sufficiently anisotropic trapping potentials, it is possible to 
reduce the condensate to a quasi-one-dimensional configuration. Typically, the 
potential is given by $V(x) = \Omega^2 x^2/2$, where the trap 
strength $\Omega \ll 1$. To a first-approximation, therefore, 
the potential can be ignored. In this limit, the condensate 
becomes homogeneous and the GP equation is identical to the 
integrable non-linear Schr\"odinger equation: 
\begin{equation}
\label{NLS}
i \hbar \partial_t \psi = -\frac{\hbar^2}{2m} 
\partial^2_x \psi + g |\psi |^2 \psi .
\end{equation}

The defocusing NLS equation ({\ref{NLS}), where $g >0$, 
supports dark soliton solutions \cite{tsu}. 
A dark soliton is an envelope excitation
characterized by a dip in the ambient density and a phase jump across 
the density minimum. Such solitons have been observed 
in a variety of Bose-Einstein condensates in recent years (for 
a recent review, see \cite{darkreview}).
The solution is given by \cite{zak}
\begin{eqnarray}
\label{soldensity}
|\psi |^2 & = & n \left( 1-\left| \psi_{\rm ds} \right|^2 
\right) 
\nonumber \\
\label{solitondensity}
\left| \psi_{\rm ds} \right|^2 & = & 
 \left( 1 - \frac{v^2}{c^2} \right) {\rm sech}^2 \, 
 \left[ \sqrt{1- \left( \frac{v}{c} \right)^2} \frac{(x-vt)}{\xi} 
\right] 
\end{eqnarray} 
where $n$ is the background density, 
$v$ is the speed of the soliton, $(x-vt)$ is its position 
and $c=\sqrt{ng/m}$ is the sound speed in 
the condensate. The spatial extent of the soliton is 
characterized by the healing length $\xi = \hbar/\sqrt{mng} = 
1/\sqrt{4 \pi n a}$. In general, the speed of the soliton is bounded from 
above by the sound  speed, $v < c$. 

The density profile of the dark soliton (\ref{solitondensity})
corresponds precisely to the energy density of the 
inflationary universe (\ref{Hsoliton}) and the LQC scaling solution 
(\ref{lqcscale}). On dimensional grounds, we 
can make the identification 
\begin{equation}
\label{dimgrounds}
\frac{\phi}{m_{\rm P}} \longleftrightarrow (x-vt)\sqrt{na}
\end{equation}
and view the scalar field as a wave-like coordinate. 
Modulo a constant of proportionality, 
the spectral index may then be identified with the speed of the soliton: 
\begin{equation}
\label{spectralspeed}
1- n_s \longleftrightarrow \left( 1- \frac{v^2}{c^2} \right) ,
\end{equation}
whereas in the LQC model, the speed is proportional 
to the kinemetic parameters:
\begin{equation}
\label{lqcbose}
\frac{\gamma}{\Omega_{\phi}} 
\longleftrightarrow \left( 1- \frac{v^2}{c^2} \right) .
\end{equation}
The deviation of the spectral index away from the scale-invariant 
perturbation spectrum is proportional to the speed of the soliton relative 
to the condensate sound speed. The maximal speed of the soliton is 
attained in the limit of the scale-invariant spectrum, $n_s \rightarrow 1$. 
For the LQC scaling solution, the maximal speed corresponds to an equation 
of state $p =-\rho$ for the fluid, which is the limit of 
a cosmological constant. 

The above analogies are not intended to be precise, but 
they do nonetheless suggest a new link 
between gravitational and non-gravitational systems might 
be established through the KdV equation. It would
be interesting to formalise such analogies further to 
establish kinematic correspondences between cosmology 
and condensed matter physics. 

\section*{Acknowledgements}

We thank R. Tavakol for numerous helpful discussions.

\end{document}